\documentclass{raa}           
\usepackage{natbib}
\usepackage{amssymb}
\usepackage{supertabular, epsfig}
\usepackage{url}
\usepackage{float}
\usepackage[caption = false]{subfig}
\usepackage{graphicx}
\usepackage{hyperref} 

\begin{document}

   \title{Binary star detection in the Open Cluster King 1 field}

\volnopage{{\bf20XX}\ {\bf X}No.{\bf XX},000--000}
   \setcounter{page}{1}

   \author{Parvej Reja Saleh\inst{1}$^{\star}$, Debasish Hazarika\inst{1}, Ajaz Ahmad Dar\inst{2}, Padmakar Singh Parihar\inst{3} and Eeshankur Saikia\inst{1}
   }

   \institute{ Department of Applied Sciences, Gauhati University-781014, Assam, India;\\
   $\star$\textit{prvzslh@gauhati.ac.in}\\
        \and
            Department of Physics, University of Kashmir-190006, Srinagar-J\&K, India
	\and
	 Indian Institute of Astrophysics, Koramagala-560034, Bangalore, India\\
\vs\no
   {\small Received~~20xx month day; accepted~~20xx~~month day}}

\abstract{A rarely studied open cluster, King 1 is observed using the $1.3m$ telescope equipped with $2k \times 4k$ CCD at the Vainu Bappu Observatory, India. We analyse the photometric data obtained from the CCD observations in both $B$ and $V$ bands. Out of 132 detected stars in the open cluster King 1 field, we have identified 4 stellar variables and 2 among them are reported as newly detected binary systems, in this paper. The parallax values from GAIA DR2 suggest that the open cluster King 1 is at the background of these two detected binary systems, falling along the same line of sight, giving rise to different parallax values. The periodogram analysis was carried out using Phase Dispersion Minimization (PDM) and Lomb Scargle (LS) method for all the detected variables. PHOEBE (PHysics Of Eclipsing BinariEs) is extensively used to model various stellar parameters of both the detected binary systems. Based on the modeling results obtained from this work, one of the binary systems is reported for the first time as an Eclipsing Detached (ED) and the other as an Eclipsing Contact (EC) binary of W-type W UMa.
}

   \authorrunning{Parvej et al.}            
   \titlerunning{Binary star detection in the King 1 field}  
   \maketitle

\section{Introduction}
\label{sect:intro}
To study the stellar evolution, the open clusters are the best objects, as they provide vital informations about their distance, age, composition and masses. Numerous types of variable stars are present in a star cluster because of their transition through various stages of evolution \citep{Lata+etal+2011}. A very rarely observed open cluster, King 1, discovered by \citet{king1949} in the second Galactic quadrant near the Perseus spiral arm, was initially investigated through photometric study in UBVRI bands by \citet{Lata2004}. Their study reports a broad main sequence in the color-magnitude diagram and a galactic radius of $4'$, derived from their density of the stellar surface. They also estimated the age, distance modulus, and reddening to be $1.6 \pm 0.4$ Gyr, $(m-M)_{0} = 11.38$ mag, and $E(B-V) = 0.70 \pm 0.05$ mag respectively, based on the isochrone fitting, considering solar metallicity. Soon after, \citet{maci2007} reported the same parameters to be $\sim4$ Gyr, $(m-M)_{0} = 10.17^{+0.32}_{-0.51}$ mag, and $E(B-V) = 0.76 \pm 0.09$ mag respectively. Also, from the $VI$ photometry of King 1, \citet{hase2008} reported the age, distance modulus, and reddening to be $2.8$ Gyr, $(m-M)_{0} = 11.57$ mag, and $E(B-V) = 0.62$ mag respectively. In \citet{Carrera+etal+2017}, the same parameters were estimated using both photometric and spectroscopic observations and was reported as $2.8 \pm 0.3$ Gyr, $(m-M)_{0} = 10.6 \pm 0.1$ mag, and $E(B-V) = 0.80 \pm 0.05$ mag respectively. Additionally, they estimated the central location of the open cluster King 1 at $RA = 00^{h}22^{m}$ and $DEC = +64^{0} 23'$ with an uncertainty of $\sim1'$. A central density, and a core radius were also computed to be $6.5 \pm 0.2$ star arcmin$^{-2}$, and $1.'9 \pm 0.'2$ respectively. Later, \citet{Lata+etal+2014} updated the age, distance, reddening, and linear diameter to be $\log$(Age) = $9.2\pm0.1$, $1.9$ kpc, $E(B - V) = 0.70\pm0.05$ mag, and $4.3$ pc respectively using $UBVRI$ CCD photomtery. However, an inconsistency in the age, distance modulus, and reddening can be clearly seen in the reported results so far.

Since, the open cluster King 1 was less studied and no previous reports were found so far with the specific goal of detecting stellar variables, therefore, our main motivation of this work was to explore the stellar variability and investigate the presence of binary stars within the field of view of the open cluster King 1. These findings  definitely add value to the existing studies of this open cluster, and also similar others. The physical characteristics of the stars, such as, mass, light, temperature and abundance play a significant role in studying the pulsating stars, as this type of stars occur on a wide range of parameters and evolutionary phases. 

This paper is organised as follows - the observational analysis and data reduction were described in the section~\ref{sect:obser}. The process of detecting the variable stars and their respective light curves are presented in the section~\ref{sect:detect}. In section~\ref{sect:period}, multiple period detection methods are performed on the variable stars. In the Section~\ref{sect:result} the results are summarized, which discusses the phase plots, Color-Magnitude diagrams, and the photometric solutions obtained from the PHOEBE (PHysics Of Eclipsing BinariEs) modeling of the detected binary stars.

\section{Photometric Observation and Data Analysis}
\label{sect:obser}
Photometric observations of the open cluster King 1 were carried out for both the B and V band at Vainu Bappu Observatory, Kavalur from November 05 to November 11, 2017, using a $1.3$ m telescope. The $2k\times{4k}$ CCD (Charged-Coupled Device) with FoV of $\sim0.3 \frac{arcsec}{pixel}$, readout noise of $\sim4.2$ electrons, and gain 0.745 $e/{ADU}$ was used to monitor the cluster. We have recorded Bias, Flat, and Object frames for each night.  The images in B and V filters were taken with long exposure times. Table~\ref{Tab1} provides the details of the observation log.

\begin{table}[htbp]
\bc
\caption[]{Observation Log of the open cluster King 1\label{Tab1}}
\begin{tabular}{c@{\hskip 0.4in}c@{\hskip 0.3in}c@{\hskip 0.3in}c} 
\hline \hline
Observation Nights & Bands & No. of Frames & Exposure (s) \\ \hline
05-11-2017& B,V  & 26 & 600, 600 \\
06-11-2017& B,V  & 09 & 600, 600 \\
07-11-2017& B,V  & 21 & 600, 600 \\
08-11-2017& B,V  & 17 & 600, 600 \\
09-11-2017& B,V  & 22 & 600, 600 \\
10-11-2017& B,V  & 27 & 600, 600 \\
11-11-2017& B,V  & 21 & 600, 600 \\ \hline \hline
\end{tabular}
\ec
\end{table}

\begin{figure}
\centering
\subfloat[132 stars]{\includegraphics[width = 2in]{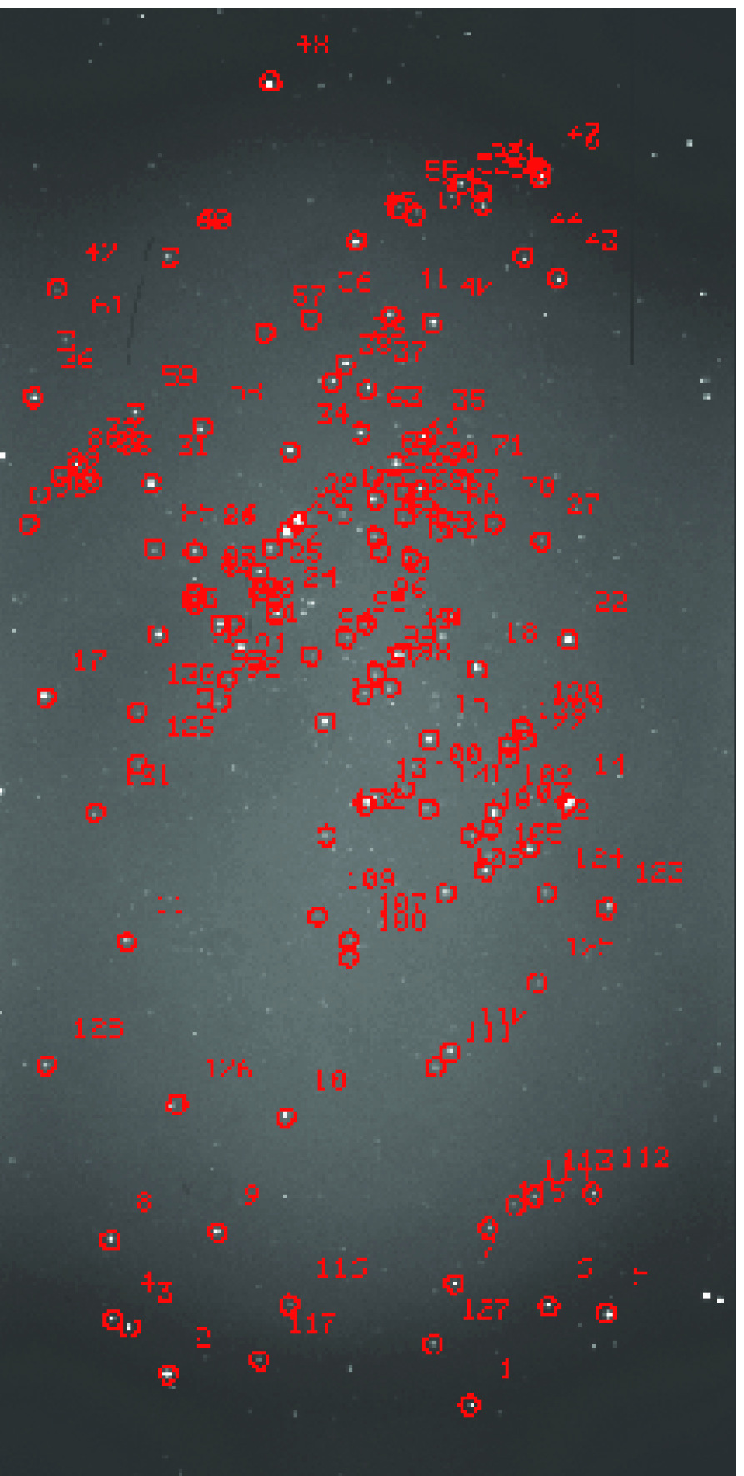}} \hspace{0.1\textwidth}
\subfloat[4 variables]{\includegraphics[width = 2in]{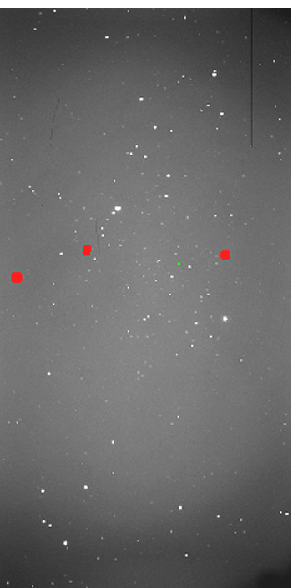}}\\
\caption{a) Photometric frames of King 1 field, marked with the detected stars, b) Among the detected stars, variable stars are marked.}
\label{Fig1}
\end{figure}

The CCD frames were cleaned and pre-processed in IRAF (Image Reduction and Analysis Facility) to recognize available bright stars in our field. By considering the image shift and the rotation, we computed the co-ordinates of the detected stars with respect to a reference frame. The reference frame was selected after a visual inspection of all the 143 object frames, which have got best photographic qualities among the others. DAOPHOT \citep{Stetson+1987} has been used to detect the stars and generate it's respective co-ordinates for every frame. The co-ordinates are used in DAOPHOT-II's PHOT task as an input to aperture photometry. The data files with night frames including all the scientific data, such as, flux, magnitude with errors of all marked stars are dumped into the DAOPHOT (PHOT) output. A maximum number of stars' fluxes were taken into consideration for computing standard deviation. The stars with lesser standard deviations were considered as comparison stars. Python programming was extensively used to develop this algorithm. Stars with larger standard deviations have been discarded. This process is iterative and is continued until the remaining stars are found to be non-variable, i.e. less standard deviations \citep{Dar+etal+2018, Dar+Padmakar+Malik+2018}. Following \citet{Kiss+etal+2001, Kang+etal+2007, Wang+etal+2015}, the instrumental magnitudes of each detected stars in all the frames were used to compute the standard magnitudes respectively. Among 132 detected stars, we have selected 4 unsaturated bright stars, having less standard deviations for differential photometry, as shown in the Figure~\ref{Fig1}. These stars are not located in any of the edges of the CCD frames. \citet{Gilland+Brown+1988}'s ensemble normalisation was used to normalize instrumental magnitudes of the V band. 

\section{Variable Star Detection}
\label{sect:detect}
\begin{figure}[htbp]
\includegraphics[width=1\textwidth]{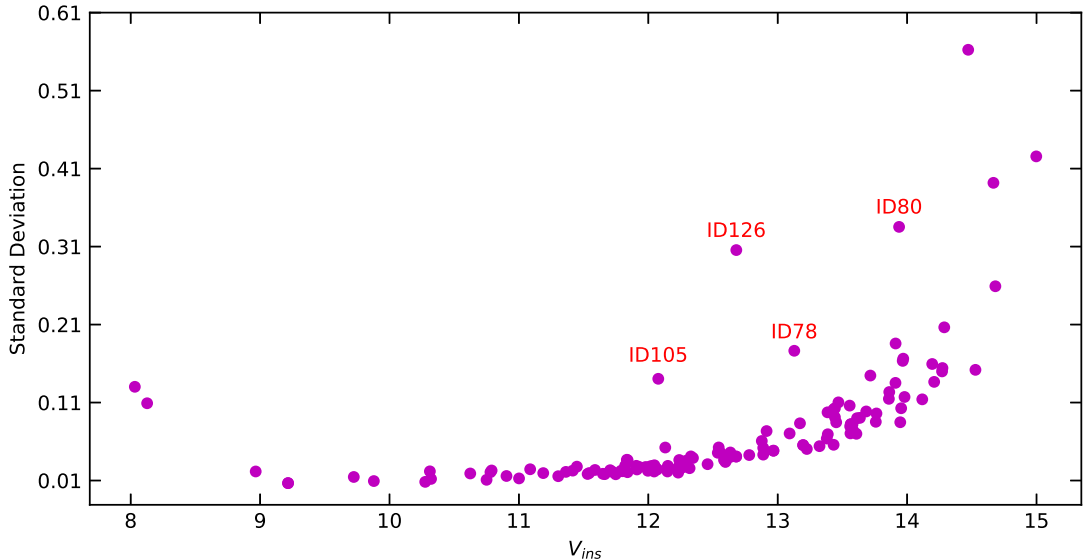}
\caption{Standard deviation vs. magnitude-averaged mean V magnitude of 132 detected stars in the  King 1 field. The marked stars are variables with relatively higher standard deviations, ignoring the outliers, which are considered for further analysis.}
\label{Fig2}
\end{figure}

The presence of variability in 132 detected stars were visually inspected from all light curves. We searched for characteristics, such as, eclipses of binaries, planetary transits and pulsations with elevated amplitude. Figure~\ref{Fig2} reflects the standard deviation in $V$ band depending on the instrumental magnitude. Out of 132 stars in the King 1 field, 4 variables were detected with higher standard deviations as compared to the other non-variant stars. These four objects have also been observed by Two Micron All-Sky Survey (2MASS)\footnote{\url{https://irsa.ipac.caltech.edu/Missions/2mass.html}}and GAIA DR2.\footnote{\url{https://www.cosmos.esa.int/web/gaia}} 2MASS reported their precise positions (RA, Dec) and brightness informations in the infrared bands i.e. J (1.25 microns), H (1.65 microns), and Ks (2.17 microns) whereas, GAIA DR2 provided the parallaxes, proper motions, radial velocity informations and G-band magnitudes \citep{gaia_dr2}. The 2MASS and GAIA DR2 IDs of the reported four variables are mentioned in Table~\ref{Tab2}.

\begin{figure}[htbp]
\subfloat[Parallax from GAIA DR2]{\includegraphics[width = 2.8in]{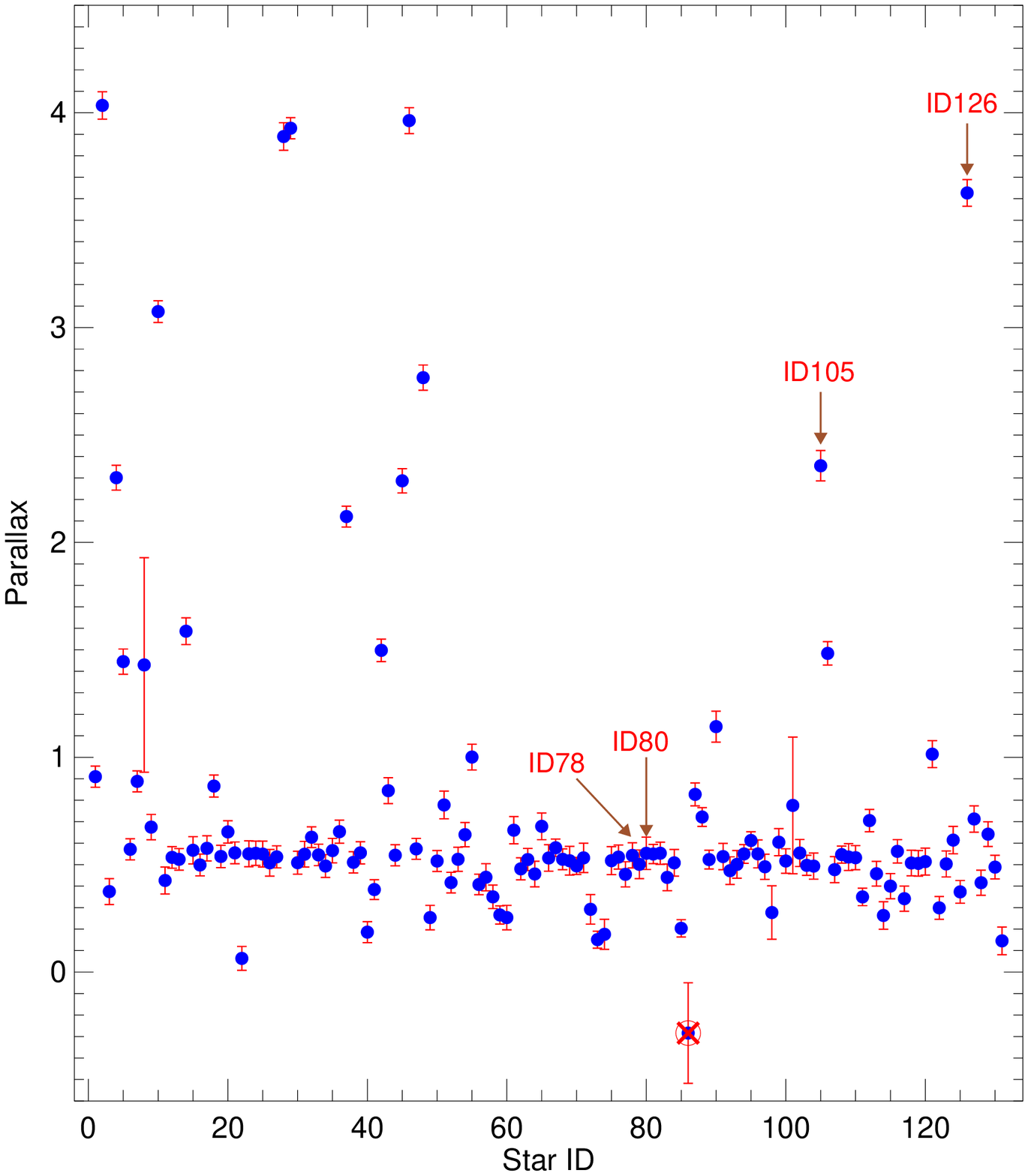}}
\subfloat[RA Dec vs Parallax]{\includegraphics[width = 2.8in]{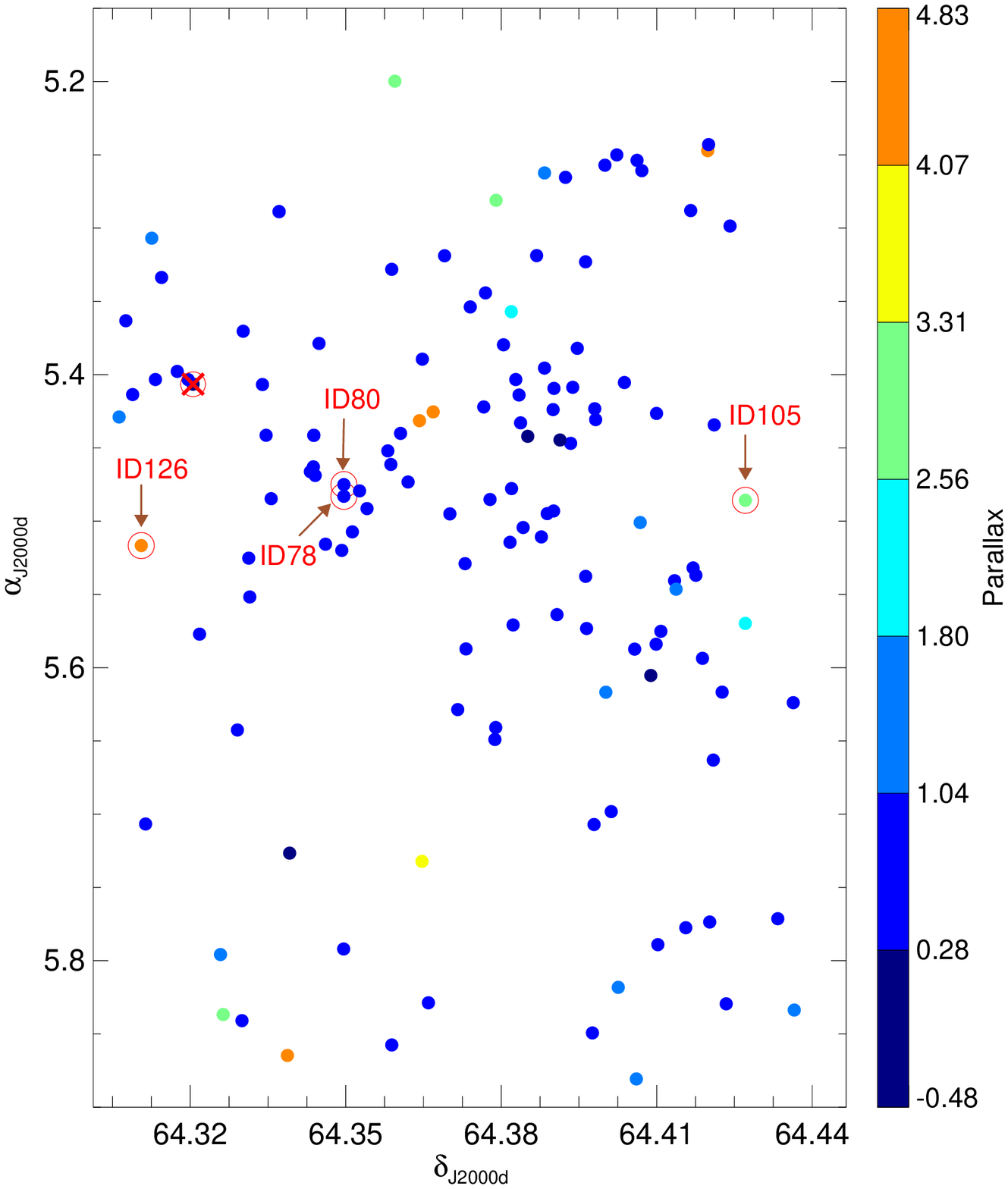}}
\caption{a) The parallax, in milliarcseconds (mas) of the 132 detected stars in the CCD frame considered along with its error is shown here. Most of the objects lie in the parallax range from 0 to 1, which includes ID78 and ID80. However, there are few objects possessing higher parallax, including ID105 and ID126. The object marked as $\bigotimes$, has a negative parallax as reported by GAIA DR2. b) The RA and Dec mapping of the 132 detected stars with their parallax values.}
\label{Fig99}
\end{figure}

The field of view of the open cluster King 1 was calibrated with the available catalogues, viz. 2MASS and GAIA DR2, where the header of one of the CCD frames was updated with the World Coordinate System (WCS). WCS in the header allows us to extract the RA and Dec of each star present in a CCD frame. To compute the WCS, a plate solution was created using the IRAF packages \textit{ccmap }and \textit{ccsetwcs}. SAOImage DS9 was extensively used to perform this process, where the GAIA DR2 catalogues \citep{gaia_dr2} was used as a reference to match the position of our CCD frame.

\begin{figure}[htbp]
\subfloat[ID078]{\includegraphics[width = 2.8in]{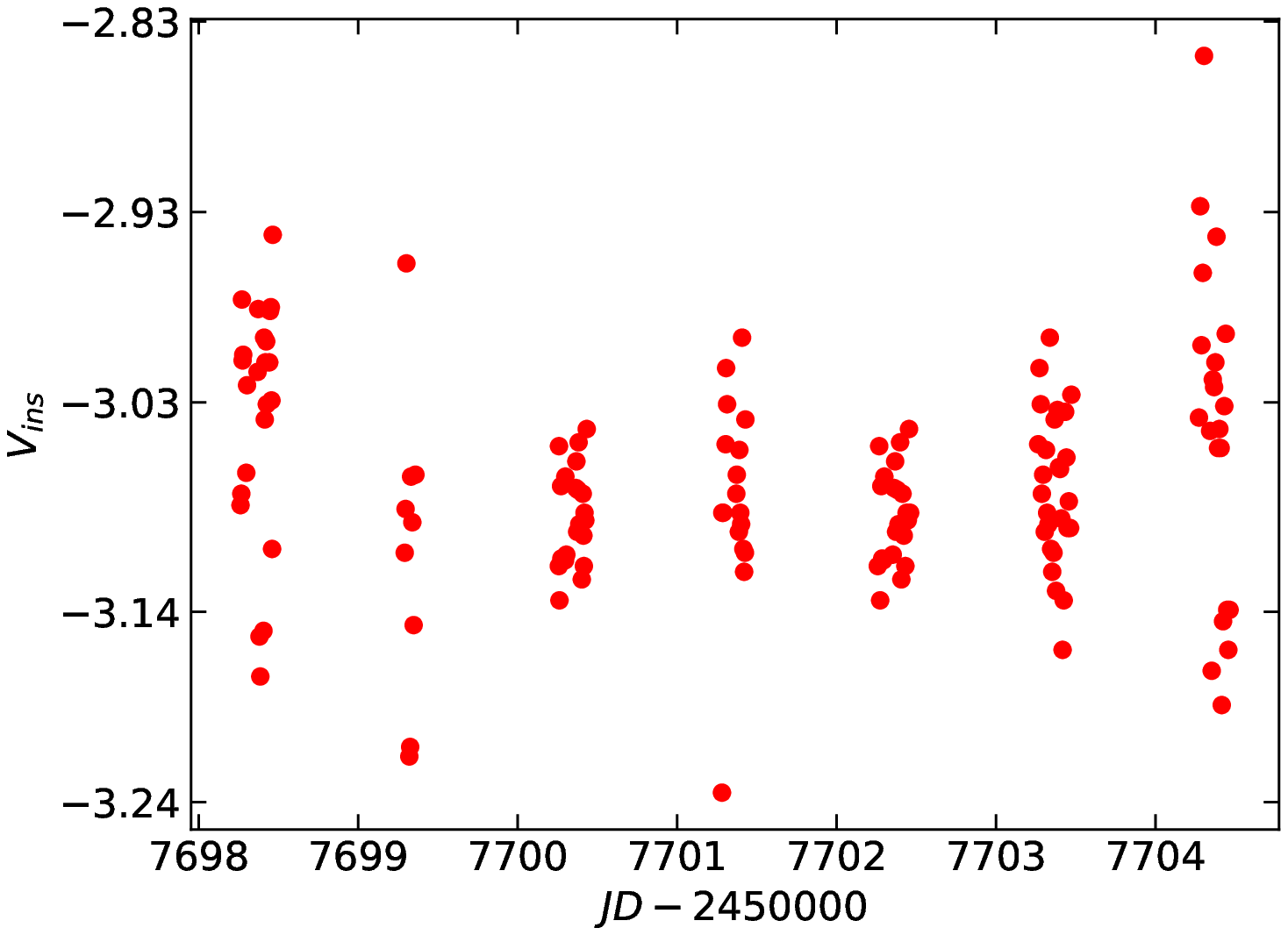}} 
\subfloat[ID080]{\includegraphics[width = 2.8in]{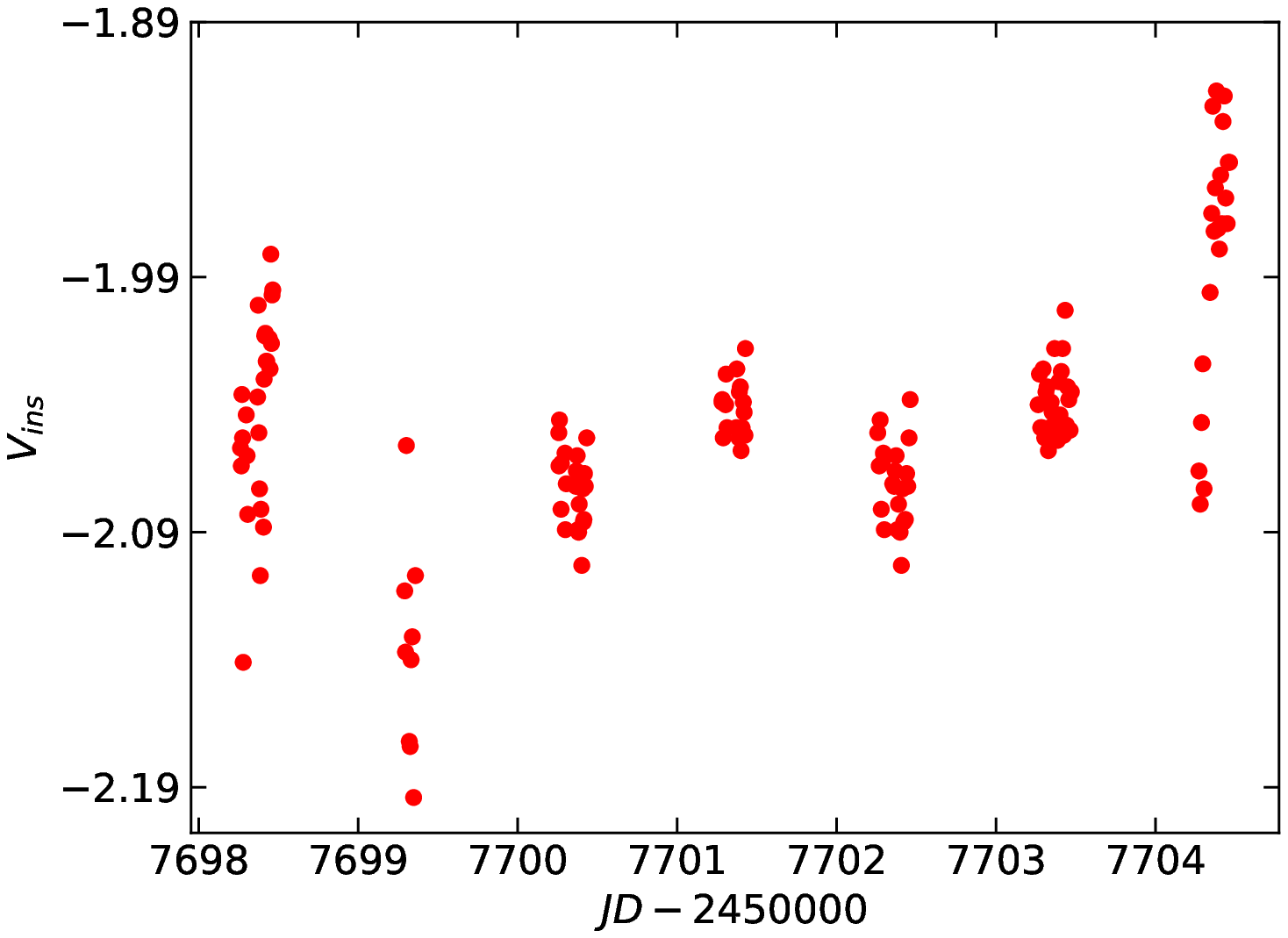}}\\[-0.1in]
\subfloat[ID105]{\includegraphics[width = 2.8in]{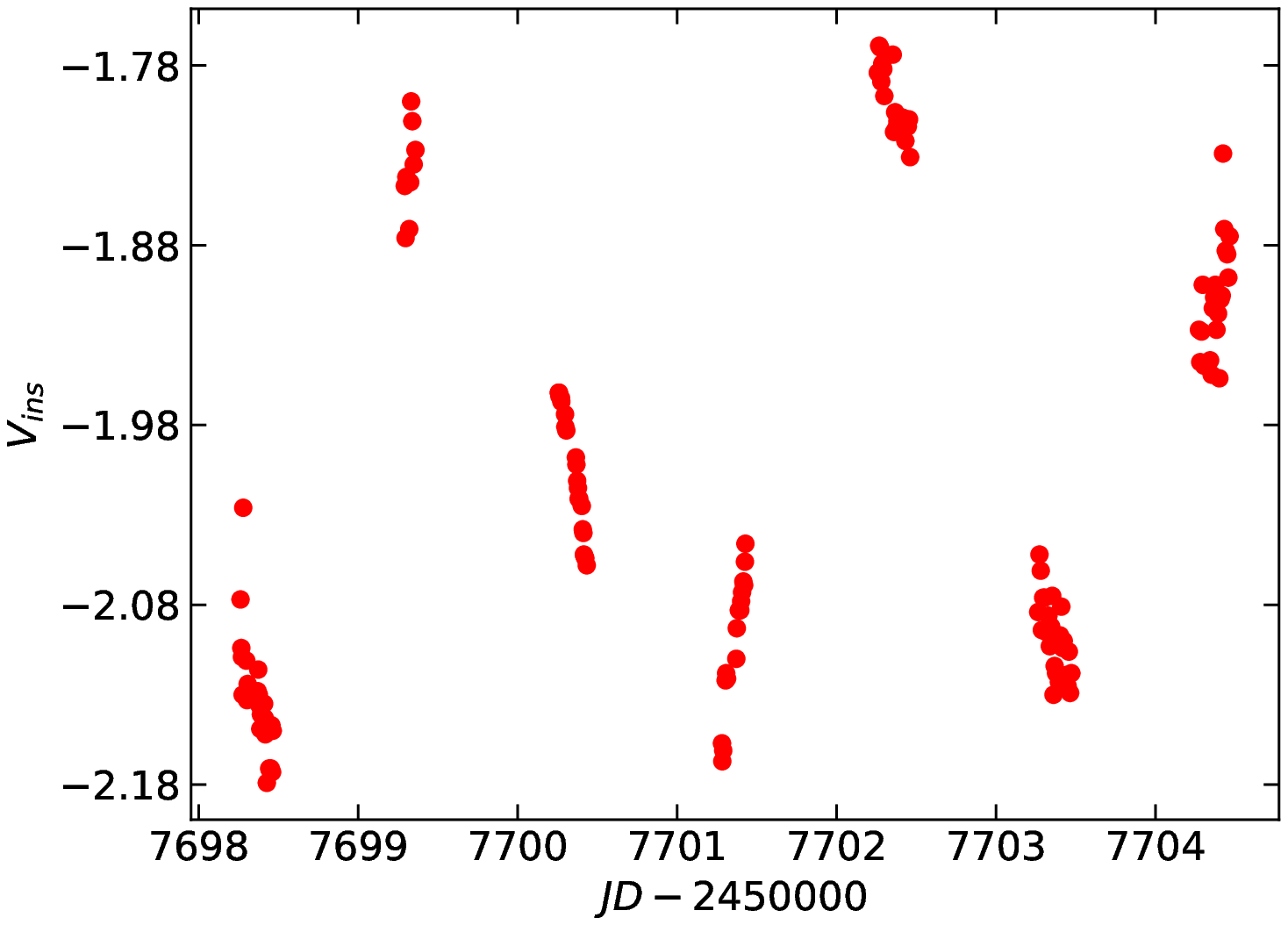}}
\subfloat[ID126]{\includegraphics[width = 2.8in]{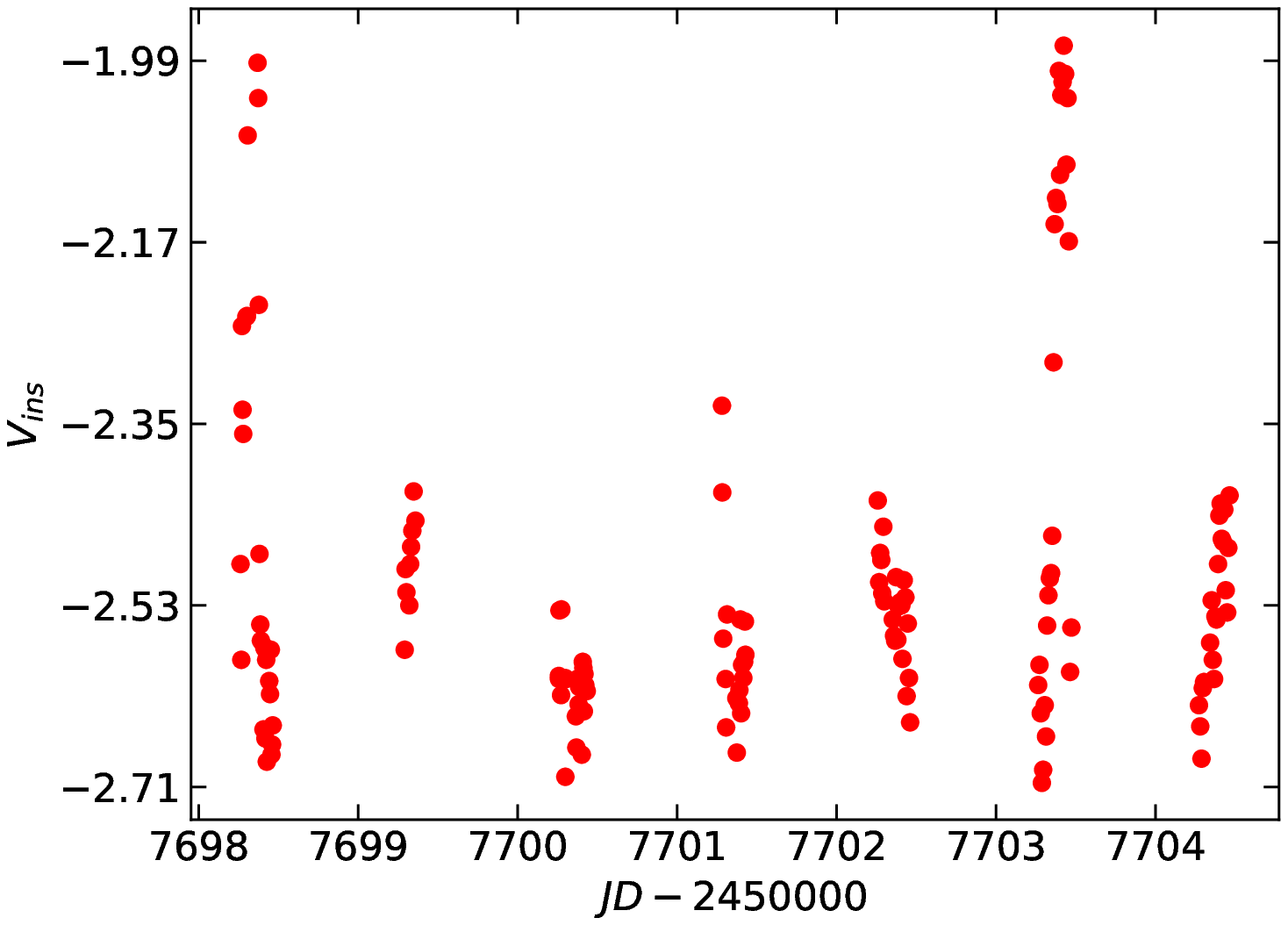}} 
\caption{Light Curves of the 4 detected variables in the  King 1 field.}
\label{Fig3}
\end{figure}

The parallax of the 132 stars present in the CCD frame are taken into account. It depicts that there are few objects which do not belong to the King 1. \citet{Lata+etal+2014} estimated the distance to the King 1 as 1.9 kpc, which corresponds to a parallax of $\sim 0.5263$ mas. Noticeably, ID78, and ID80 have a parallax of 0.5430$\pm$0.0285 mas and 0.5529$\pm$0.0755 mas respectively, supporting their existence within the open cluster. However, the two reported binaries ID105 and ID126, with a parallax of 2.3568$\pm$0.0706 mas and 3.6268$\pm$0.0621 mas respectively, seem to exist much ahead of the King 1, though along its line of sight. Nevertheless, we have extended our work on these two binary stars, as no previous discussion on their characteristics are reported.

\section{Period Determination}
\label{sect:period}

To obtain the most probable periods, we have used two methods, viz. Phase Dispersion Minimization (PDM) \citep{Stellingwerf+1987}, and Lomb Scargle periodogram (LS) \citep{Scargle+1982}. PDM and LS are extremely fast and precise method in determining the most significant period in an unevenly sampled data of a variable star \citep{Horne+Baliunus+1986}. We have performed periodogram analysis on the four detected variables, covering a frequency range between 0.05 and 100 day$^{-1}$, with a maximum of 1000 frequency steps in PDM. 

The light curves of the 4 detected variables are shown in Figure~\ref{Fig3}. These are visually inspected and phase plots are generated for each of the variable star, based on the periods obtained from PDM method.

\begin{table}[htbp]
\centering
\scriptsize
\caption[]{Period of the variables identified in King 1\label{Tab2}}
\resizebox{\textwidth}{!}{
\begin{tabular}{cccccccc}
\hline \hline
Object    & RA   & Dec   & Period	& $\log$(P)	& Period     		& ID	& ID \\
 & (hh:mm:ss) & (dd:mm:ss) & (PDM)	& (PDM)		& (LS)			& (GAIA DR2)   & (2MASS) \\ \hline
ID078 &  00:21:55.93 &  +64:20:58.70 & 1.1976	& 0.2957  & 1.1853	& 431180635962447104    & 00215593+6420587 \\
ID080 &  00:21:54.81 &  +64:20:58.60 & 1.5325	& 0.1854  & 1.6559	& 431180635954092288   & 00215481+6420586  \\
ID105 &  00:22:03.92 &  +64:18:37.53 & 5.3150	& 0.7255  & 5.393	& 431177264404482944   & 00220394+6418380  \\
ID126 &  00:21:56.53 &  +64:25:37.89 & 0.7253   & -0.1390 & 0.7336	& 431182422668794880  & 00215654+6425381   \\ \hline \hline
\end{tabular}}
\end{table}

We are not reporting the phase plots of ID078 and ID080, since their phase plots point towards aperiodic variations. For ID105 and ID126, the PDM based period gives less scattered and well arranged phase plot as compared to the periods obtained from LS. The determination of periods is found to improve further by using the entropy minimization method, as described in \citet{Deb+etal+2010}. The entropy minimization (ME) method is run around $\pm 1 \%$  of the periods obtained from PDM. However, both the periods from PDM and ME methods are found to be nearly identical in the present case. Table~\ref{Tab2} contains the detected periods of the four variables, along with their catalogue IDs and RA Dec.

\section{Results and Conclusions}
\label{sect:result}
In eclipsing binaries, a portion of light is blocked when one component moves over the other one and the flux is diminished. They display particular characteristics and are readily acknowledged by observational data. Vital information, such as, radius, mass ratio, inclination angles, temperatures, etc. can be obtained for these types of star by modeling their variation of light with respect to different phases of eclipses. 

\begin{figure}[htbp]
\subfloat[ID105 ($P=5.31500$ days)]{\includegraphics[width = 2.8in]{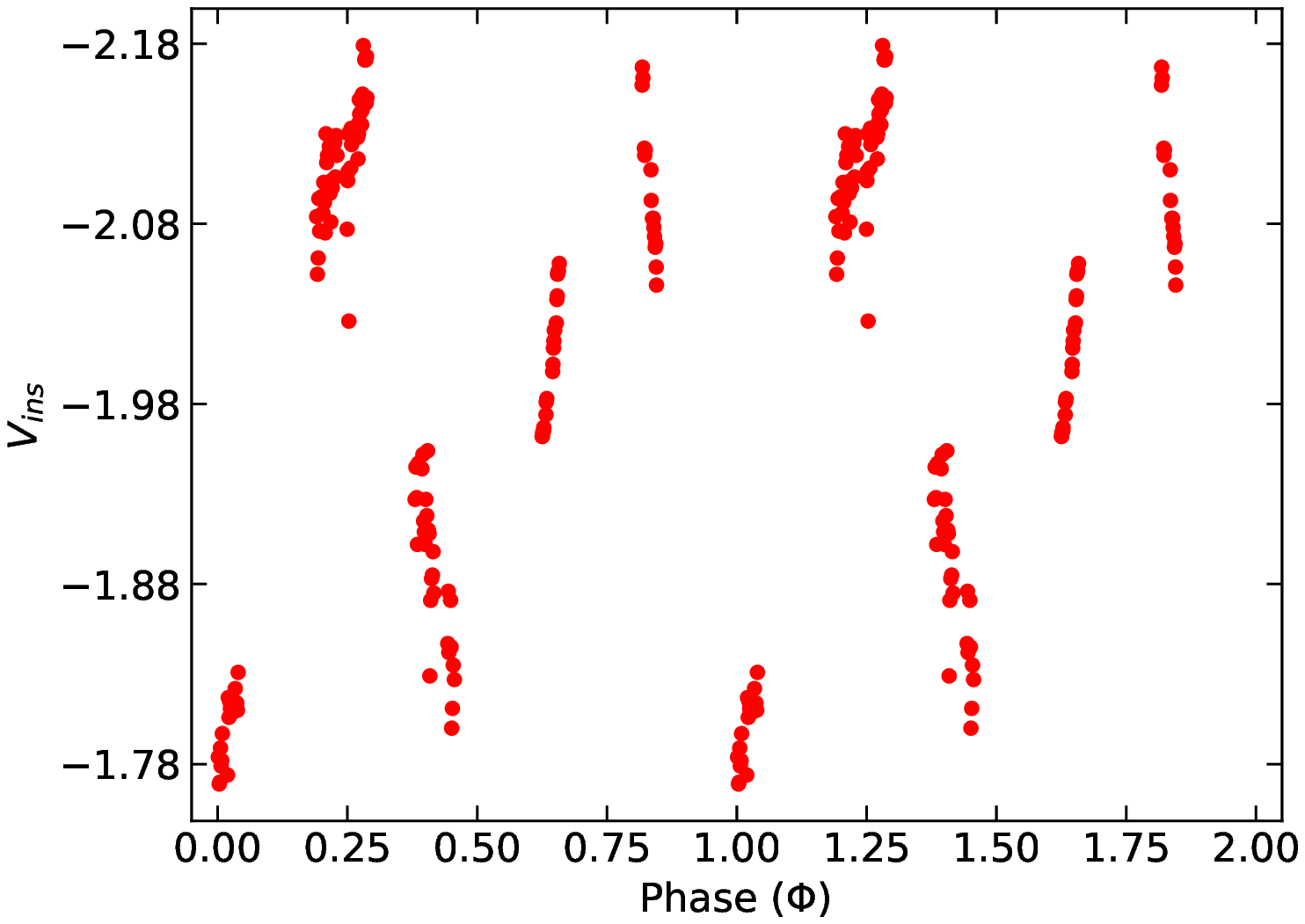}} 
\subfloat[ID126 ($P=0.72538$ days)]{\includegraphics[width = 2.8in]{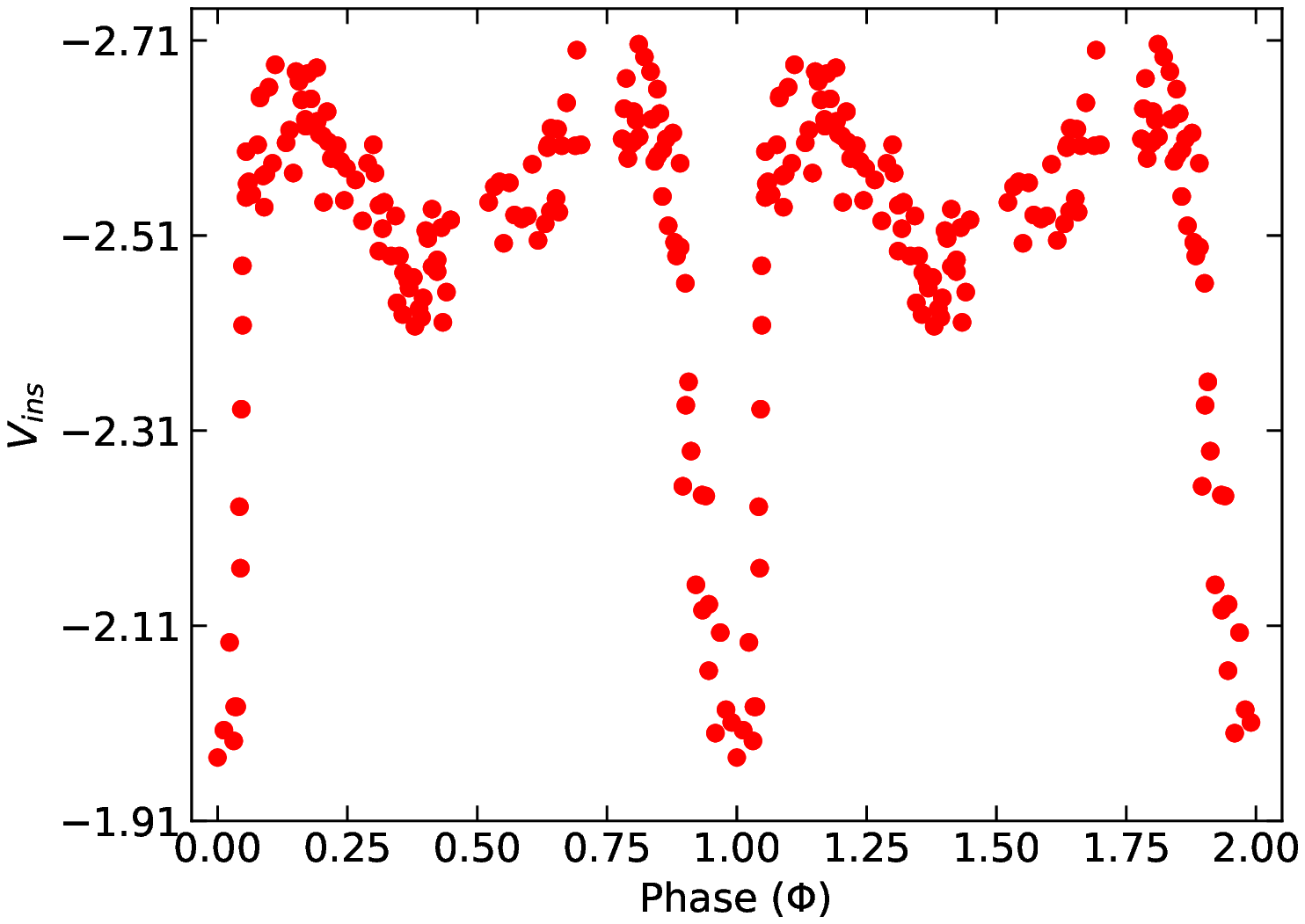}}
\caption{Phased light curves of the binary stars (ID105 \& ID126) of the King 1 field. The period value obtained from PDM and ME method is used to construct the phased plots.}
\label{Fig4}
\end{figure}

From the photometric observation of King 1 field, we detected 4 stellar variables, out of which 2 are found to be binary stars. However, the parallax of these four variables obtained from GAIA DR2 confirms that ID105 and ID126 are not a part of King 1, and located much ahead of the cluster, though along its line of sight. Figure~\ref{Fig4} shows the phase plot of the  detected binaries. To analyse the observed light curves, we have used PHOEBE based on W-D code \citep{Wilson+Devinney+1971}.  In PHOEBE, the parameters, such as, gravity brightening, limb darkening, and bolometric albedo are kept as default for both the stars. For convective envelopes, the gravity brightening coefficients for binaries are 0.32, whereas the bolometric albedo is taken as 0.6 for both the components. Table~\ref{Tab3} shows the photometric solutions obtained for the two detected binaries. The fitted light curves obtained from PHOEBE modeling are shown in Figure~\ref{Fig6}. We have followed \citet{Lucy+etal+1967}, while applying PHOEBE for the over-contact binaries.

\begin{figure}[htbp]
\subfloat[ID105]{\includegraphics[width = 2.8in]{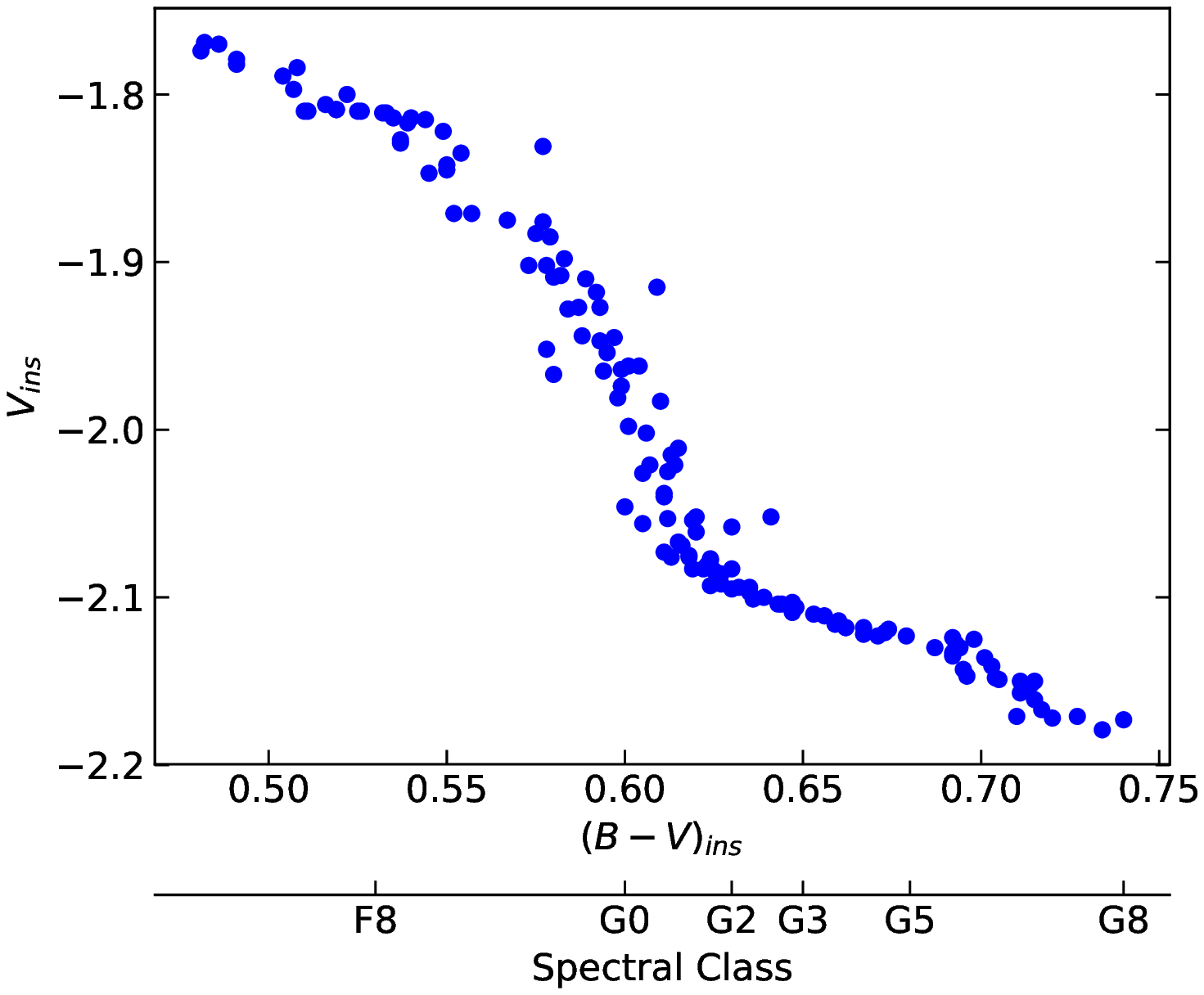}} 
\subfloat[ID126]{\includegraphics[width = 2.8in]{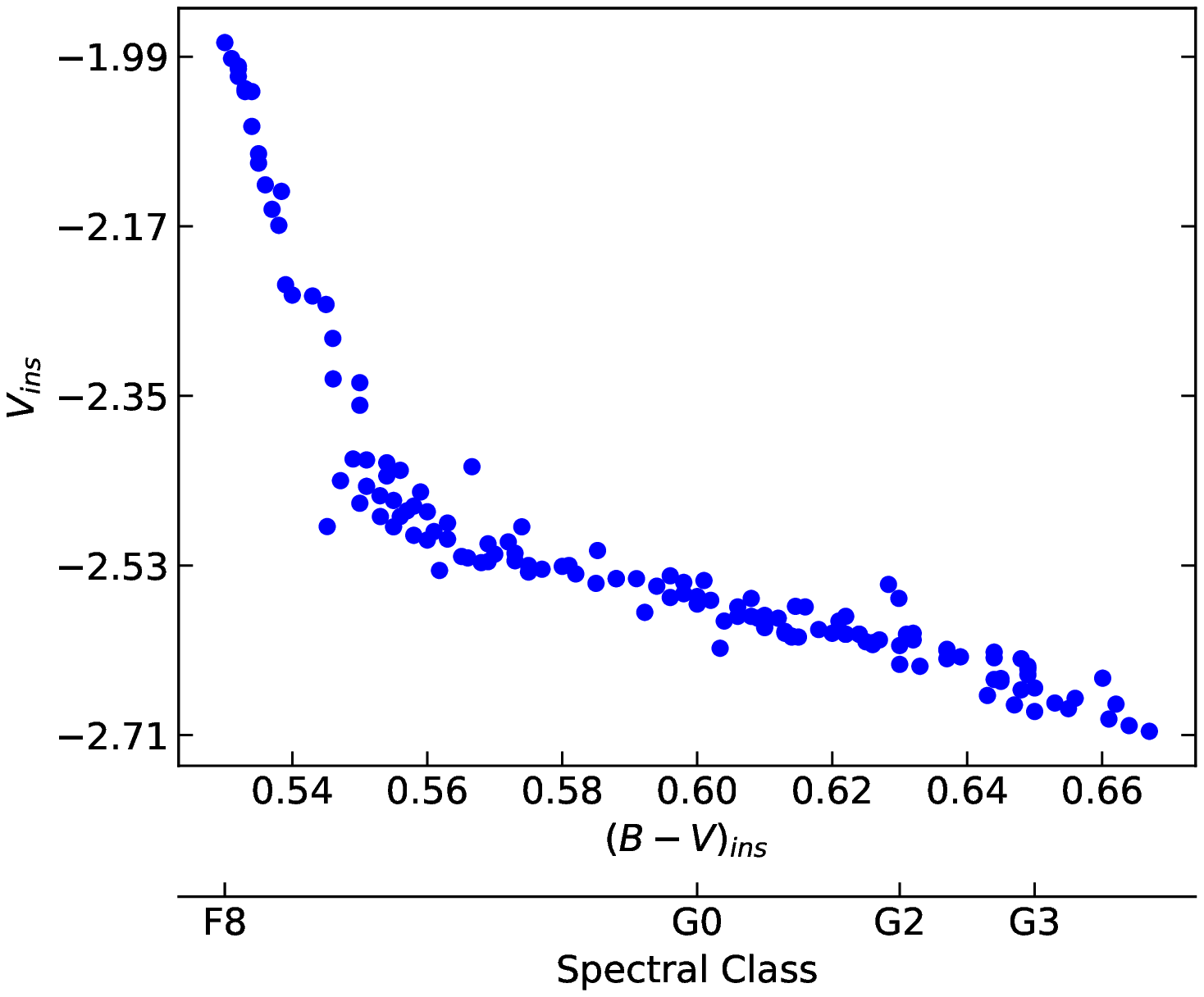}}
\caption{Color-Magnitude Diagrams of both the binary stars in the open cluster King 1 field. The spectral class axes in both (a) and (b) relate the Color (B-V) values to their corresponding Spectral Class.}
\label{Fig5}
\end{figure}

\begin{figure}[htbp]
\subfloat[ID105]{\includegraphics[width = 2.8in]{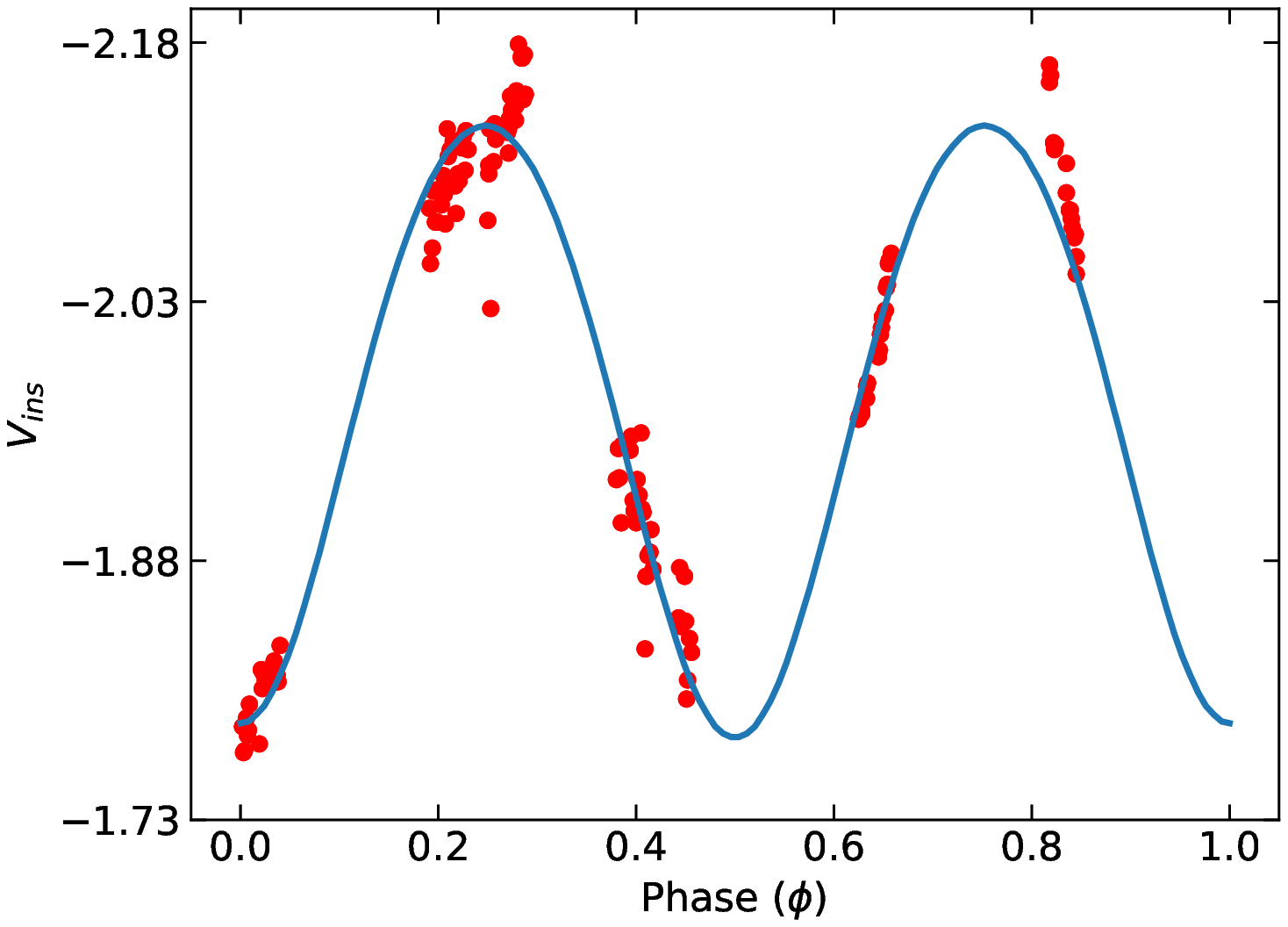}} 
\subfloat[ID126]{\includegraphics[width = 2.8in]{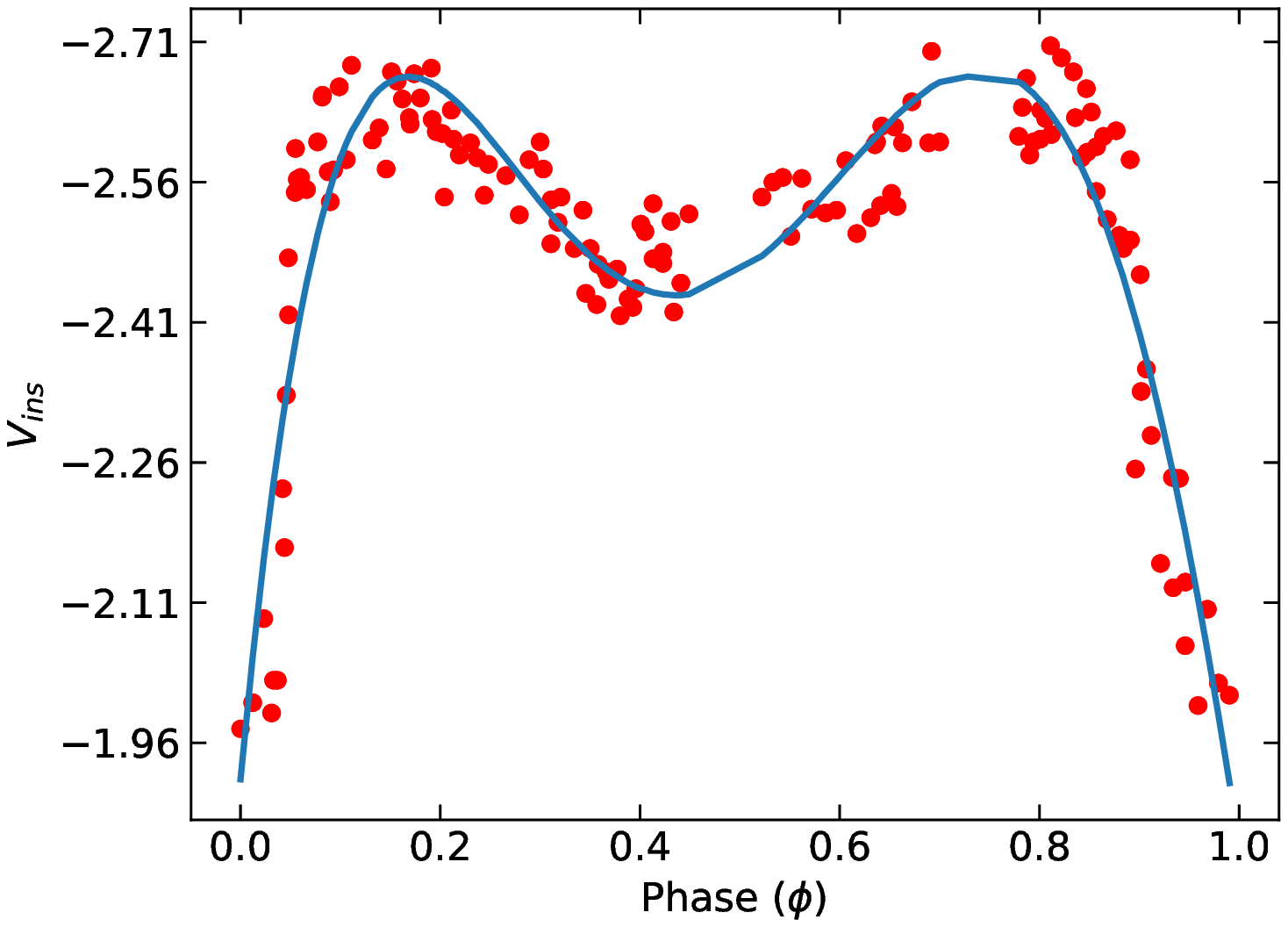}}
\caption{PHOEBE modeling of the binary stars (ID105 \& ID126) in the King 1 field.}
\label{Fig6}
\end{figure}

Mass ratio is a significant parameter for understanding a binary star. W UMa binaries usually have short periods of in the range $0.2 < p < 1$ day \citep{Kang+etal+2002, Lau+etal+2017}. Based on observational features W UMa binaries can be divided into two subclasses, i.e. i) A-type ii) W-type. The A-type system with greater luminosity and reduced mass ratio i.e $q=M_2/M_1<{1}$ are of spectral type from A to G \citep{Binnendijk+1970}. The transit eclipse of the massive hot component shows deeper minima for a W UMa binary of A-type. The W-type W UMa binary generally holds a mass ratio, $q=M_2/M_1>{1}$ and a spectral type from F to K. The angle of inclination ($i$) of the stellar systems ID105 and ID126, are found to be $51.8^\circ$ and $57.5^\circ$ respectively. Stars with inclination $i<90^{\circ}$ has counter clock-wise rotation, as mentioned in \citet{Dar+Saleh+2018}. The temperature difference $\Delta{T}$ between the components of ID105 and ID126 are $475K$ and $122K$ respectively. For difference in temperatures between the components $\Delta T \geq 1000 K$, the binary system is classified as B-type W UMa, where the component stars are not in thermal contact, but in geometrical contact with each other \citep{Rucinski+Duerbeck+1997, Deb+etal+2010}.
However, from our analysis, we did not find W UMa binary of B-type, as the temperature difference of both the components is found to be less than 1000K for ID105 and ID126 respectively.

On the other hand, the period of the binary star ID105 is found to be 5.393 days, which is greater than 1 day, confirming that the binary system cannot be a W UMa. \citet{li2008} classified the three types of binary systems, i.e. Contact (EC), Semi Detached (ESD), and Detached (ED), using the logarithm of Period ($\log$ P) values of the variable stars, where most of the EC, ESD, and ED systems ranges around $\sim -0.455$, $\sim -0.2$, and $\sim 0.35$ respectively. In our analysis, ID105 and ID126 holds $\log$ P values $ 0.7255$ and $-0.139$ respectively, as mentioned in Table~\ref{Tab2}. Considering the $\log$ P value and mass ratio of ID105 and ID126, we can conclude that, ID105 is an Eclipsing Detached Binary (ED), whereas ID126 is an Eclipsing Contact (EC) binary.

\begin{table}[htbp]
\bc
\footnotesize
\begin{minipage}[]{140mm}
\caption{Physical parameters of the binary stars in the King 1 field obtained from PHOEBE\label{Tab3}}\end{minipage}
\resizebox{\textwidth}{!}{
\begin{tabular}{l@{\hskip 0.08in}c@{\hskip 0.07in}c@{\hskip 0.06in}c@{\hskip 0.06in}c@{\hskip 0.07in}c@{\hskip 0.07in}c@{\hskip 0.07in}c@{\hskip 0.07in}c@{\hskip 0.07in}c@{\hskip 0.07in}c@{\hskip 0.07in}c@{\hskip 0.05in}c@{\hskip 0.07in}c@{\hskip 0.07in}c@{\hskip 0.07in}c@{\hskip 0.05in}c}
\hline \hline
Star  & \textit{i} & $\Omega_1$ & $\Omega_2$ & $q=\frac{m_{2}}{m_{1}}$ & $T_{1}^{e}$ & $T_{2}^{e}$ & $R_1$ & $R_2$ & $M_{2}^{b}$ & $M_{1}^{b}$ & $\log(g)_{1}$ & $\log(g)_{2}$ & SB$_{1}$ & SB$_{2}$ & PHSV & PCSV\\ \hline
ID105 & 51.8                        & 2.987      & 2.657      & 0.558       & 5725              & 6200              & 4.676 & 3.827 & 1.476       & 1.565       & 3.896        & 3.818        & 0.048           & 0.070           & 4.3  & 9    \\
ID126 & 57.5                        & 10.922     & 10.277     & 6.445       & 5578              & 5700              & 1.224 & 4.418 & 4.499       & 1.619       & 4.112        & 3.807        & 3.982           & 0.346           & 12.5 & 10   \\ \hline \hline
\end{tabular}}
\ec
$\Omega_{1,2}$: Luminosity of [1] primary \& [2] secondary star,$\quad T_{1,2}^{e}$: Effective Temperature in K,$\quad$ SB$_{1,2}$: Surface Brightness,$\quad$ PHSV: Primary Star Surface Potential,$\quad$ PCSV: Secondary Star Surface Potential.
\end{table}

The period, mass ratio, and effective temperature of ID126 crowns it to be a newly detected W UMa binary system. It's mass ratio being greater than 1 and the spectral type between F to  G suggest it to be a W type W UMa binary star. The effective temperatures obtained from PHOEBE modeling ranges between 5500 K to 6000 K for both the binary systems, as mentioned in Table~\ref{Tab3}. The range of effective temperature is in a good agreement with the results obtained from the Color-Magnitude Diagram, as shown in Figure~\ref{Fig5}. The Color-Magnitude diagram in Figure~\ref{Fig5}, defends the spectral types of both the binary systems, correlating the effective temperatures calculated from the photometric observations with their modeling counterpart obtained from PHOEBE.

We have reported the photometric study of two variables, viz. ID78 and ID80 from the King 1 cluster and two binary systems, viz. ID105 and ID126 in the King 1 field, along the line of sight of the open cluster. Though the results reported are definitely going to add values to the understanding of this relatively less studied cluster, coupling spectroscopic as well as polarimetric results with it would enhance our insight into the field.

\section*{Acknowledgement}
\label{sect:acknow}

PRS and DH is thankful to MHRD TEQIP-III for awarding fellowship for pursuing Ph.D. at Gauhati University. PRS and AAD is also grateful to the Director, Indian Institute of Astrophysics (IIA), Bangalore for providing observation time at Vainu Bappu Observatory. A hearty thanks to the anonymous referee for carefully examining the manuscript, helping us to explore new dimensions in the study and adding values to the work.  

This work has made use of data from the European Space Agency (ESA) mission {\it Gaia} (\url{https://www.cosmos.esa.int/gaia}), processed by the {\it Gaia} Data Processing and Analysis Consortium (DPAC, \url{https://www.cosmos.esa.int/web/gaia/dpac/consortium}). Funding for the DPAC has been provided by national institutions, in particular the institutions participating in the {\it Gaia} Multilateral Agreement. 

\label{bibliography}
\bibliographystyle{apalike}
\bibliography{raa-2019-0247_r4} 
\label{lastpage}
\end{document}